\documentclass{article}
\usepackage{ijcai20}

\usepackage{times}
\usepackage{soul}
\usepackage{url}
\usepackage[hidelinks]{hyperref}
\usepackage[utf8]{inputenc}
\usepackage[small]{caption}
\usepackage{graphicx}
\usepackage{amsmath}
\usepackage{amsthm}
\usepackage{amssymb}
\usepackage{booktabs}
\usepackage{algorithm}
\usepackage{algorithmic}
\usepackage{subfigure}
\usepackage{amsfonts}
\usepackage{color}

\newtheorem{definition}{Definition}

\title{Improving Knowledge Tracing via Pre-training Question Embeddings}

\author{
Yunfei Liu$^1$\and
Yang Yang$^1$\and
Xianyu Chen$^1$\and
Jian Shen$^1$\and
Haifeng Zhang$^2$\And
Yong Yu$^1$\footnote{Corresponding author.}\\
\affiliations
$^1$Shanghai Jiao Tong University\\
$^2$The Center on Frontiers of Computing Studies, Peking University\\
\emails
\{liuyunfei, yyang0324, xianyujun\}@sjtu.edu.cn,
rockyshen@apex.sjtu.edu.cn,\\
pkuzhf@pku.edu.cn,
yyu@apex.sjtu.edu.cn
}

\begin{document}

\maketitle

\begin{abstract}
Knowledge tracing (KT) defines the task of predicting whether students can correctly answer questions based on their historical response. Although much research has been devoted to exploiting the question information, plentiful advanced information among questions and skills hasn't been well extracted, making it challenging for previous work to perform adequately. In this paper, we demonstrate that large gains on KT can be realized by pre-training embeddings for each question on abundant side information, followed by training deep KT models on the obtained embeddings. To be specific, the side information includes question difficulty and three kinds of relations contained in a bipartite graph between questions and skills. To pre-train the question embeddings, we propose to use product-based neural networks to recover the side information. As a result, adopting the pre-trained embeddings in existing deep KT models significantly outperforms state-of-the-art baselines on three common KT datasets.


\end{abstract}
\section{Introduction}
The computer-aided education (CAE) systems are seeking to use advanced computer-based technology to improve students' learning ability and teachers' teaching efficiency \cite{cingi2013computer}. Knowledge tracing (KT) is an essential task in CAE systems, which aims at evaluating students' knowledge state over time based on their learning history. To be specific, the objective of KT is to predict whether a student can answer the next question correctly according to all the previous response records.

To solve KT problem, various approaches have been proposed including Bayesian Knowledge Tracing (BKT) \cite{corbett1994knowledge,zhu2018integrating}, the factor analysis models \cite{wilson2016back,pavlik2009performance} and deep models \cite{piech2015deep,zhang2017dynamic}. In this paper, we mainly focus on the deep KT models, which leverage recent advances in deep learning and have achieved great success in KT. In general, most deep KT models estimate a student's mastery of skills instead of directly predicting her capability to answer specific questions correctly. Two representative methods are DKT \cite{piech2015deep} and DKVMN \cite{zhang2017dynamic}. 

\begin{figure}
    \centering
    \includegraphics[width=0.45\textwidth]{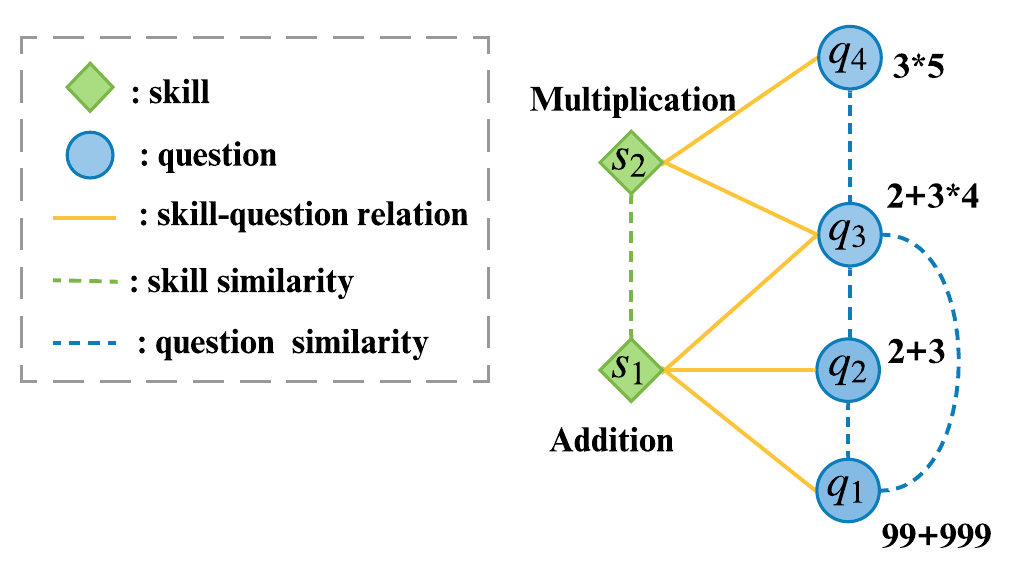}
    \caption{Illustration of a question-skill bipartite graph. The question-skill relations are the explicit relations, and the skill similarity and question similarity are implicit relations. Questions $q_1$ and $q_2$ share the same skill $s_1$ but have different difficulties so that skill-level mastery modeling is insufficient. But the implicit similarity between $q_1$ and $q_2$ can help prediction to tackle the sparsity issue.}
    \label{fig:1}
\end{figure}

Although skill-level mastery can be well predicted by these deep KT models, there exists a major limitation that the information of specific questions is not taken into consideration \cite{piech2015deep,zhang2017dynamic,Ghodai2019Knowledge}. As shown in Figure \ref{fig:1}, the questions sharing the same skill may have different difficulties, and thus skill-level prediction can not accurately reflect the knowledge state of a student for specific questions. Although it is quite necessary to solve KT at a finer-grained level by exploiting the information of specific questions, there comes a major issue that the interactions between students and questions are extremely sparse, which leads to catastrophic failure if directly using questions as the network input \cite{DBLP:conf/edm/WangMG19}. 
To tackle the sparsity issue, several works are proposed to use the question information as a supplement \cite{minn2019dynamic,DBLP:conf/edm/WangMG19}. However, these works only consider the question difficulties or question-skill relations.



In this paper, we take a further step towards maximally extracting and exploiting plentiful underlying information among questions and skills to tackle the sparsity issue. Considering that usually a skill includes many questions and a question is also associated with several skills, we can represent them as a bipartite graph where vertices are skills and questions respectively. Generally, bipartite graphs include two kinds of relations \cite{gao2018bine}: the explicit relations (\textit{i.e.}, observed links) and the implicit relations (\textit{i.e.}, unobserved but transitive links). In KT scenarios as shown in Figure \ref{fig:1}, in addition to the explicit question-skill relations, we consider the implicit skill similarity and question similarity, which haven't been well exploited in previous work.


Taking everything into consideration, in this paper, we propose a pre-training approach, called Pre-training Embeddings via Bipartite Graph (PEBG), to learn a low-dimensional embedding for each question with all the useful side information. To be specific, the side information includes question difficulties together with three kinds of relations: explicit question-skill relations, implicit question similarity and skill similarity. To effectively extract the knowledge contained in the side information, we adopt a product layer to fuse question vertex features, skill vertex features and attribute features to produce our final question embeddings. In this way, the learned question embeddings will preserve question difficulty information and the relations among questions and skills. 

The contributions of this paper are summarized as follows.
\begin{itemize}
\item To the best of ours, we are the first to use the bipartite graph of question-skill relations to obtain question embeddings, which provides plentiful relation information.
\item We propose a pre-training approach called PEBG, which introduces a product layer to fuse all the input features, to obtain the final question embeddings.
\item The obtained question embeddings by PEBG can be incorporated into existing deep KT models. Experiment results on three real-world datasets show that using PEBG can outperform the state-of-the-art models, improving AUC by $8.6\%$ on average.
\end{itemize}
\section{Related Work}

Previous KT methods can be largely categorized into three types: Bayesian Knowledge Tracing (BKT), factor analysis KT models and deep KT models. 
\cite{corbett1994knowledge} proposes the Bayesian Knowledge Tracing (BKT) model, which is a hidden Markov model and assumes students' knowledge state as a set of binary variables. BKT models each skill state separately, making it unable to capture the relations among skills.

Another line of KT methods is factor analysis, which considers the factors that affect student state, including the difficulty of questions, students' ability, the ratio of correct answers to a certain question. The factor analysis models include Item Response Theory (IRT) \cite{wilson2016back}, Additive Factor Model (AFM) \cite{cen2006learning}, Performance Factor Analysis (PFA) \cite{pavlik2009performance}, Knowledge Tracing Machine (KTM) \cite{vie2019knowledge}. 
These models only consider the historical interactions of each question or skill, and also fail to capture the relations between questions and skills.

With the rise of deep learning, lots of deep models have been proposed to solve KT, among which most preliminary work uses skills as network input. For example, \cite{piech2015deep} proposes the Deep Knowledge Tracing (DKT) model, which uses a recurrent neural network (RNN) to model the learning process of students. Dynamic Key-Value Memory Network (DKVMN), proposed by \cite{zhang2017dynamic}, uses a key-value memory network to automatically discover the relations between exercises and their underlying concepts and traces each concept state. 
The PDKT-C model \cite{chen2018prerequisite} manually labels the prerequisite relations among skills, 
which however is not suitable for large-scale data. The GKT model \cite{nakagawa2019graph} builds a similarity graph of skills randomly, and automatically learns the edge weights of the graph to help prediction.

Since the skill-level prediction cannot fully reflect the knowledge state of specific questions, several works propose to use the question information as a supplement. For example,  \cite{su2018exercise,huang2019ekt} encode text descriptions of questions into question embeddings to capture the question characteristics, but the text descriptions are not easy to acquire in practice. \cite{minn2019dynamic} calculates the percentage of incorrect answers as the question difficulty to distinguish different questions. DHKT \cite{DBLP:conf/edm/WangMG19} uses relations between questions and skills as a constraint to train question embeddings, which are used as the input of DKT together with skill embeddings. In this paper, we mainly focus on how to pre-train a low-dimensional embedding for each question, which can be directly used as the network input. 

\section{Problem Formulation}

In knowledge tracing, given a student's past question interactions $\mathcal{X}=\{ (q_1,c_1),..., (q_{t-1}, c_{t-1}) \}$ where $c_i$ is the correctness of the student's answer to the question $q_i$ at the time step $i$, the goal is to predict the probability that the student will correctly answer a new question, \textit{i.e.}, $P(c_t=1|q_t, \mathcal{X})$.

Let $Q=\{ q_i \}_{i=1}^{|Q|}$ be the set of all distinct $|Q|$ questions and $S=\{ s_j \}_{j=1}^{|S|}$ be the set of all distinct $|S|$ skills. Usually, one skill includes many questions and one question is related to several skills.
So the question-skill relations can be naturally represented as a bipartite graph $G=(Q, S,\pmb{R})$ where $\pmb{R}=[r_{ij}] \in \{0,1\}^{|Q|\times |S|}$ is a binary adjacency matrix. If there is an edge between the question $q_i$ and the skill $s_j$, then $r_{ij}=1$; otherwise $r_{ij}=0$. Here we introduce the information we will use to train embeddings in our model, including the information in the graph and the difficulty information.



\begin{definition}[explicit question-skill relations]
Given the question-skill bipartite graph, relations between skill vertices and question vertices are the explicit question-skill relations, that is, explicit relation between question vertex i and skill vertex j depends on whether $r_{ij}$ =1.
\end{definition} 


\begin{definition}[implicit question similarity and skill similarity]
Given the question-skill bipartite graph, relations between two skill vertices that have the common neighbor question vertices are defined as skill similarity. Similarly, question similarity refers to the relations between two question vertices that share the common neighbor skill vertices.
\end{definition} 

    

\begin{definition}[question difficulty] The question difficulty $d_i$ for one question $q_i$ is defined as the ratio of correctly being answered computed from the training dataset. All the question difficulties form a vector $\pmb{ d}=[d_i] \in \mathbb{R}^{|Q|}$.
\end{definition}


\section{Method}

\begin{figure*}[ht]
    \centering
    \includegraphics[width=0.95\textwidth]{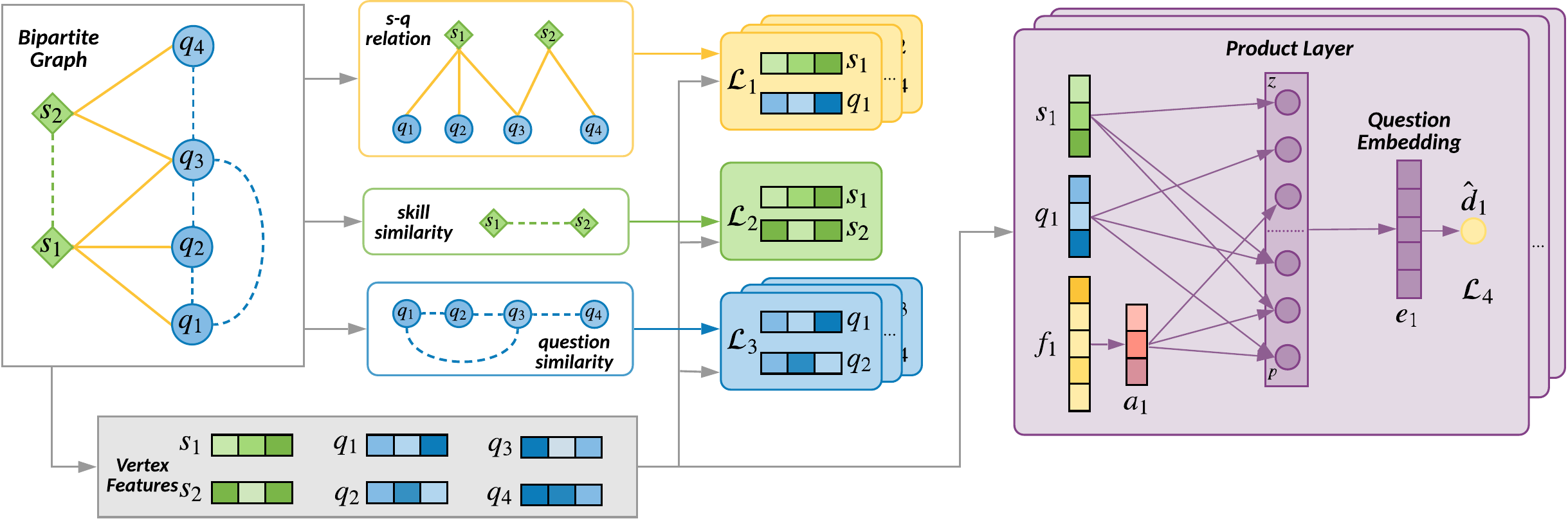}
    \caption{The PEBG framework overview.}
    \label{fig:framework}
\end{figure*}

In this section, we will give a detailed introduction of our PEBG framework, of which the overview architecture is given by Figure \ref{fig:framework}. PEBG pre-trains question embeddings using four loss functions respectively designed for the side information: explicit skill-question relations, implicit question similarity and skill similarity, and question difficulty.



\subsection{Input Features}

To pre-train the question embeddings, we use three kinds of features as follows. It should be noted that the vertex features are initialized randomly and will be updated in the pre-training stage, which is equivalent to learning linear mappings from one-hot encodings to continuous features.

\textbf{Skill vertex features} are represented by a feature matrix $\pmb{S}\in \mathbb{R}^{|S|\times d_v}$, where $d_v$ is the dimension of the features. For one skill $s_i$, the vertex feature is denoted as $\pmb{s}_i$, which is the $i$-th row of matrix $\pmb{S}$.

\textbf{Question vertex features} are represented by a feature matrix $\pmb{Q}\in \mathbb{R}^{|Q|\times d_v}$, which has the same dimension $d_v$ as the skill vertex features. For one question $q_j$, the vertex feature is denoted as $\pmb{q}_j$, which is the $j$-th row of matrix  $\pmb{Q}$.

\textbf{Attribute features} are the features related to the difficulty of questions, such as average response time, question type and so on. For question $q_i$, we concatenate the features as $\pmb{f}_i=[\pmb{f}_{i1};..;\pmb{f}_{im}]$, $m$ is the number of features. $\pmb{f}_{ij}$ is a one-hot vector if the $j$-th feature is categorical (e.g., question type). $\pmb{f}_{ij}$ is a scalar value if the $j$-th feature is numerical (e.g., average response time).


\subsection{Bipartite Graph Constraints}

The skill and question vertex features are updated via the bipartite graph constraints.
As there exist different relations in the graph, we design different types of constraints so that the vertex features can preserve these relations.


\subsubsection{Explicit Question-Skill Relations}
In the question-skill bipartite graph, edges exist between question vertices and skill vertices, presenting an explicit signal. Similarly to the modeling of 1st-order proximity in LINE \cite{tang2015line}, we model explicit relations by considering the local proximity between skill and question vertices.
In detail, we use inner products to estimate the local proximity between question and skill vertices in the embedding space,
\begin{equation}
    \hat{r}_{ij} = \sigma(\pmb{ q}_i^T\pmb{ s}_j),i \in [1,...,|Q|],j \in [1,...,|S|],
\end{equation}
where $\sigma(x)=1/(1+e^{-x})$ is the sigmoid function, which transforms the relation value to a probability.

To preserve the explicit relations, the local proximity is enforced to be close to skill-question relations in the bipartite graph via a cross-entropy loss function:
\begin{equation}
    \mathcal{L}_1(\pmb{Q},\pmb{S}) = \sum_{i=1}^{|Q|}\sum_{j=1}^{|S|} -(r_{ij} log \hat{r}_{ij} + (1-r_{ij})log (1-\hat{r}_{ij})).
\end{equation}



\subsubsection{Implicit Similarities}

The implicit similarities used in PEBG indicate the similarity between neighborhoods in the bipartite graph. Specifically, there exist two kinds of similarities: skill similarity and question similarity. We would like to use implicit similarities to update the vertex features simultaneously.

 
We define the neighbor set of question $q_i$ as $\Gamma_Q(i)=\{s_j | r_{ij}=1\}$, and the neighbor set of skill $s_j$ as $\Gamma_S(j)=\{q_i | r_{ij}=1\}$. Then the question similarity matrix $\pmb{R}^Q = [r^q_{ij}] \in \{0,1\}^{|Q|\times|Q|}$ can be formally defined as,
\begin{equation}
    r^q_{ij} = 
    \begin{cases}
        1& \Gamma_Q(i) \cap \Gamma_Q(j)\neq\emptyset\\
        0& \text{otherwise}
    \end{cases},i,j\in [1,...,|Q|].
\end{equation}
Similarly, we define the skill similarity matrix $\pmb{R}^S = [r^s_{ij}] \in \{0,1\}^{|S|\times|S|}$ as,
\begin{equation}
    r^s_{ij} = 
    \begin{cases}
        1& \Gamma_S(i) \cap \Gamma_S(j)\neq\emptyset\\
        0& \text{otherwise}
    \end{cases},i,j\in [1,...,|S|].
\end{equation}

We also use inner products to estimate the implicit relations among questions and skills in the vertex feature space,
\begin{align}
    \hat{r}^q_{ij} &= \sigma(\pmb{ q}_i^T\pmb{ q}_j),i,j\in [1,...,|Q|],\\
    \hat{r}^s_{ij} &= \sigma(\pmb{ s}_i^T\pmb{ s}_j),i,j\in [1,...,|S|].
\end{align}

We minimize the cross entropy to make vertex features preserve the implicit relations:
\begin{align}
    \mathcal{L}_2(\pmb{Q}) = \sum_{i=1}^{|Q|}\sum_{j=1}^{|Q|} -(r^q_{ij} log \hat{r}^q_{ij} + (1-r^q_{ij})log (1-\hat{r}^q_{ij})),
    \\
    \mathcal{L}_3(\pmb{S}) = \sum_{i=1}^{|S|}\sum_{j=1}^{|S|} -(r^s_{ij} log \hat{r}^s_{ij} + (1-r^s_{ij})log (1-\hat{r}^s_{ij})).
\end{align}

\subsection{Difficulty Constraint}


Difficulty information of questions is important in KT prediction, which, however, is not contained in the bipartite graph. Thus we hope the final question embeddings can recover the difficulty information. \cite{vie2019knowledge} use Factorization Machines \cite{rendle2010factorization} to  encode side information and explore feature interactions for student modeling. In this paper, we use attribute features interacting with vertex features to learn high quality embeddings. Especially, inspired by PNN \cite{qu2016product}, a product layer is used to learn high-order feature interactions.


For one question $q$ (its subscript is omitted for clarity), we have its question vertex feature $\pmb{q}$ and its attribute features $\pmb{f}$. To interact attribute features with the vertex features via a product layer, we first use a linear layer parameterized by $\pmb{w}_a$ to map the attribute features $\pmb{f}$ to a low-dimensional feature representation, which is denoted as $\pmb{a}\in \mathbb{R}^{d_v}$. Assume the set of skills related to $q$ is $C=\{s_j\}_{j
=1}^{|C|}$, we use the average representation of all skill vertex features in $C$ as the related skill feature of $q$, denoted as $\pmb{ s}'$. Mathematically,
\begin{equation}
    \pmb{ s}' = \frac{1}{|C|} \sum_{s_j\in C} \pmb{ s}_j.
\end{equation}

We use vertex feature $\pmb{ q}$, the average skill feature $\pmb{ s}'$, and the attribute features $\pmb{ a}$ to generate the linear information $\pmb{Z}$ and the quadratic information $\pmb{P}$ for the question $q$. Specifically,
\begin{align}
    \pmb{Z} &= (\pmb{ z}_1, \pmb{ z}_2, \pmb{ z}_3) \triangleq (\pmb{ q}, \pmb{s}', \pmb{ a}), \\
    \pmb{P} &= [p_{ij}] \in \mathbb{R}^{3\times 3},
\end{align}
where $p_{ij}= g(\pmb{z}_i, \pmb{z}_j)$ defines the pairwise feature interaction. There are different implementations for $g$. In this paper, we define $g$ as vector inner product: $g(\pmb{z}_i, \pmb{z}_j) =<\pmb{z}_i, \pmb{z}_j>$.

Then we introduce a product layer, which can transform these two information matrices to signal vectors $\pmb{l}_z$ and $\pmb{l}_p$, as shown in Figure \ref{fig:framework}. The transformation equations are as follows: 

\begin{align}
    l_z^{(k)} &= \pmb{W}^{(k)}_z \odot \pmb{Z} = \sum_{i=1}^{3}\sum_{j=1}^{d_v} (w^{(k)}_z)_{ij} z_{ij}, \\
    l_p^{(k)} &=\pmb{W}^{(k)}_p \odot \pmb{P} = \sum_{i=1}^{3}\sum_{j=1}^{3} (w^{(k)}_p)_{ij} p_{ij}.  
\end{align}
$k\in [1,...d]$.
And $\odot$ denotes operations that firstly element-wise multiplication is applied to two matrices, then the multiplication result is summed up to a scalar. $d$ is the transform dimension of $\pmb{l}_z$ and $\pmb{l}_p$. $\pmb{W}^{(k)}_z$ and $\pmb{W}^{(k)}_p$ are the weights in the product layer.

According to the definition of $\pmb{P}$ and the commutative law in vector inner product, $\pmb{P}$ and $\pmb{W}^{(k)}_p$ should be symmetric, so we can use matrix factorization to reduce complexity. By introducing the assumption that $\pmb{W}^{(k)}_p = \pmb{ \theta}^{(k)}\pmb{ \theta}^{(k)^T}$ and $\pmb{ \theta}^{(k)} \in \mathbb{R}^3$, we can simplify the formulation of $l_p^{(k)}$ as,
\begin{equation}
   \pmb{W}^{(k)}_p \odot \pmb{P} = \sum_{i=1}^{3}\sum_{j=1}^{3} \theta^{(k)}_i \theta_j^{(k)} <\pmb{ z}_i, \pmb{ z}_j>.
\end{equation}

Then, we can calculate the embedding of question $q$, which is denoted as $\pmb{e}$:
\begin{equation}
    \pmb{ e} = \text{ReLU}(\pmb{ l}_z + \pmb{ l}_p + \pmb{ b}),
\end{equation}
where $\pmb{ l}_z$, $\pmb{ l}_p$ and the bias vector $\pmb{ b} \in \mathbb{R}^{d}$, and $\pmb{ l}_z = (l_z^{(1)}, l_z^{(2)},...l_z^{(d)})$, $\pmb{ l}_p=(l_p^{(1)}, l_p^{(2)},...l_p^{(d)})$. The activation function is rectified linear unit (ReLU), defined as ReLU(x) = max(0, x).

To preserve the difficulty information effectively, for one question $q_i$, we use a linear layer to map the activation $\pmb{e}_i$ to a difficulty approximation $\hat{d}_i=\pmb{ w}_d^T\pmb{e}_i+b_d$ where $\pmb{w}_d$ and $b_d$ are network parameters. We use the question difficulty $d_i$ as the auxiliary target, and design the following loss function $\mathcal{L}_4$ to measure the difficulty approximation error:
\begin{align}
    \mathcal{L}_4(\pmb{Q},\pmb{S}, \pmb{\theta}) &= \sum_{i=1}^{|Q|} (d_i - \hat{d}_i)^2,
\end{align}
where $\pmb{\theta}$ denotes all the parameters in the network, \textit{i.e.}, $\pmb{\theta}=\{ \pmb{w}_a, \pmb{W}_z, \pmb{W}_p, \pmb{w}_d, \pmb{b},b_d  \}$.

\subsection{Joint Optimization}
To generate question embeddings that preserve explicit relations, implicit similarities, and question difficulty information simultaneously, we combine all the loss functions to form a joint optimization framework, namely, we solve:
\begin{equation}
    \text{min}_{\pmb{Q},\pmb{S}, \pmb{\theta}} ~ \lambda(\mathcal{L}_1(\pmb{Q},\pmb{S})+\mathcal{L}_2(\pmb{Q})+\mathcal{L}_3(\pmb{S}))+(1-\lambda)\mathcal{L}_4(\pmb{Q},\pmb{S}, \pmb{\theta}),
    \label{eq:optimization}
\end{equation}
where $\lambda$ is a coefficient to control the trade-off between bipartite graph constraints and difficulty constraint. 

Once the joint optimization is finished, we can obtain the question embeddings $\pmb{e}$, which can be used as the input of existing deep KT models, such as DKT and DKVMN. 
\section{Experiments}

In this section, we conduct experiments to evaluate the performance of knowledge tracing models based on the question embeddings pre-trained by our proposed model PEBG\footnote{Experiment code: \url{https://github.com/lyf-1/PEBG}}.

\subsection{Datasets}
We use three real-world datasets, and the statistics of the three datasets are shown in Table \ref{tab:dataset}.

\textbf{ASSIST09}\footnote{\url{https://sites.google.com/site/assistmentsdata/home/assistment-2009-2010-data/skill-builder-data-2009-2010}} and \textbf{ASSIST12}\footnote{\url{https://sites.google.com/site/assistmentsdata/home/2012-13-school-data-with-affect}} are both collected from the ASSISTments online tutoring platform \cite{feng2009addressing}. For both datasets, we remove records without skills and scaffolding problems. We also remove users with less than three records. After preprocessing, ASSIST09 dataset consists of 123 skills, 15,911 questions answered by 3,841 students which gives a total number of 190,320 records. ASSIST12 dataset contains 265 skills, 47,104 questions answered by 27,405 students with 1,867,167 records.

\textbf{EdNet}\footnote{\url{https://github.com/riiid/ednet}} is collected by \cite{choi2019ednet}. In this experiment, we use EdNet-KT1 dataset which consists of students' question-solving logs, and randomly sample 222,141 records of 5,000 students, with 13,169 questions and 188 skills.

\begin{table}
\centering
\begin{tabular}{cccc}
\hline
  & ASSIST09 & ASSIST12 & EdNet \\
\hline
\#students & 3,841 & 27,405 & 5,000   \\
\#questions & 15,911 & 47,104 & 13,169      \\
\#skills & 123 & 265 & 188 \\
\#records & 190,320 & 1,867,167 & 222,141   \\
questions per skill & 156 & 177 & 149    \\
skills per question & 1.207 & 1.000 & 2.276 \\
attempts per question & 11 & 39 & 17 \\
attempts per skill & 1,139 & 7,045 & 1,165      \\
\hline
\end{tabular}
\caption{Dataset statistics.}
\label{tab:dataset}
\end{table}

\subsection{Compared Models}
To illustrate the effectiveness of our model and show the improvement of our model to the existing deep KT models, we compare prediction performance among state-of-the-art deep KT models. We divide the compared models as skill-level models and question-level models.
\subsubsection{Skill-level Models} 
Skill-level models only use skill embeddings as input, and they all trace students' mastery of skills. 
\begin{itemize}
\item \textbf{BKT} \cite{corbett1994knowledge} is a 2-state dynamic Bayesian network, defined by initial knowledge, learning rate, slip and guess parameters.
\item \textbf{DKT} \cite{piech2015deep} uses recurrent neural network to model student skill learning.
\item \textbf{DKVMN} \cite{zhang2017dynamic} uses a key-value memory network to store the skills' underlying concept representations and states.

\end{itemize}

\subsubsection{Question-level Models}
Besides skill-level models, the following models utilize question information for question-level prediction. 
\begin{itemize}
\item \textbf{KTM} \cite{vie2019knowledge} utilizes factorization machines to make prediction, which lets student id, skill id, question features interact with each other.
\item \textbf{DKT-Q} is our extension to the DKT model, which directly uses questions as the input of DKT and predicts students' response for each question.
\item \textbf{DKVMN-Q} is our extension to the DKVMN model, which directly uses questions as the input of DKVMN and predicts students' response for each question.
\item \textbf{DHKT} \cite{DBLP:conf/edm/WangMG19} is the extension model of DKT, which models skill-question relation and can also predict students' response for each question.
\end{itemize}

We test our model based on skill-level deep learning models. \textbf{PEBG+DKT} and \textbf{PEBG+DKVMN} utilize question embeddings pre-trained by PEBG and make DKT and DKVMN achieve question-level prediction.


\subsection{Implementation Details}
To evaluate the performance of each dataset, we use the area under the curve (AUC) as an evaluation metric. 

PEBG has only a few hyperparameters.
The dimension of vertex features $d_v$ is set to 64. The final question embeddings dimension $d=128$. $\lambda$ in Eqn.(\ref{eq:optimization}) is 0.5. We use the Adam algorithm to optimize our model, and mini-batch size for three datasets is set to 256, the learning rate is 0.001. We also use dropout with a probability of 0.5 to alleviate overfitting. We divide each dataset into $80\%$ for training and validation, and $20\%$ for testing. For each dataset, the training process is repeated five times, we report the average test AUC.

For ASSIST09 and ASSIST12 datasets, average response time and question type are used as attribute features. For the EdNet dataset, average response time is used as an attribute feature.

\subsection{Performance Prediction}
Table \ref{tab:auc} illustrates prediction performance for all compared models, we find several observations as below.

The proposed PEBG+DKT and PEBG+DKVMN models achieve the highest AUC on all three datasets. Particularly, on the ASSIST09 dataset, our PEBG+DKT and PEBG+DKVMN models achieve an AUC of 0.8287 and 0.8299, which represents a significant gain of $9.18\%$ on average in comparison with 0.7356 and 0.7394 achieved by DKT and DKVMN. On the ASSIST12 dataset, the results show an average increase of $8\%$,  AUC 0.7665 in PEBG+DKT and 0.7701 in PEBG+DKVMN compared with AUC 0.7013 in DKT and 0.6752 in DKVMN. On the EdNet dataset, PEBG+DKT and PEBG+DKVMN achieve an average improvement of $8.6\%$ over the original DKT and DKVMN. 

Among all the compared models, BKT has the worst performance. DKT, DKVMN, and KTM have similar performance. By comparing the performance of DKT and DKT-Q, DKVMN and DKVMN-Q, we find DKT-Q and DKVMN-Q show no advantage, which indicates that directly applying existing deep KT models to question-level prediction will suffer from question interactions sparsity issue. And our PEBG model can improve DKT and DKVMN well, even on those sparse datasets. 
Though DHKT outperforms DKT, it still performs worse than our proposed model, which illustrates the effectiveness of PEBG in leveraging more complex relations among skills and questions.

\begin{table}
\centering
\begin{tabular}{cccc}
\hline
Model & ASSIST09 & ASSIST12 & EdNet\\
\hline
BKT & 0.6476 & 0.6159 & 0.5621 \\
DKT & 0.7356 & 0.7013 & 0.6909   \\
DKVMN & 0.7394 & 0.6752 & 0.6893      \\
\hline
KTM & 0.7500  & 0.6948 & 0.6855 \\
DKT-Q & 0.7244 & 0.6899 & 0.6876   \\
DKVMN-Q & 0.7405 & 0.6812 & 0.7152  \\
DHKT & 0.7544 & 0.7213 & 0.7245 \\
\hline
PEBG+DKT & 0.8287 & 0.7665 & \textbf{0.7765}  \\
PEBG+DKVMN & \textbf{0.8299} & \textbf{0.7701} & 0.7757    \\
\hline
\end{tabular}
\caption{The AUC results over three datasets.}
\label{tab:auc}
\end{table}

\begin{table}
\centering
\begin{tabular}{cccc}
\hline
Model & ASSIST09 & ASSIST12 & EdNet\\
\hline
RER+DKT & 0.8144 & 0.7584 &    0.7652  \\
RER+DKVMN & 0.8053 & 0.7617    & 0.7663  \\
\hline
RIS+DKT & 0.8082 & 0.7608 & 0.7622 \\
RIS+DKVMN & 0.8063 & 0.7603 & 0.7657 \\
\hline
RPL+DKT & 0.7763 & 0.7355 & 0.7445   \\
RPL+DKVMN & 0.7623 & 0.7033 & 0.7437  \\
\hline
RPF+DKT & 0.8151 & 0.7473 & 0.7528   \\
RPF+DKVMN & 0.8127 & 0.7391 & 0.7533  \\
\hline
PEBG+DKT & 0.8287 & 0.7665 & \textbf{0.7765}  \\
PEBG+DKVMN & \textbf{0.8299} & \textbf{0.7701} & 0.7757    \\
\hline
\end{tabular}
\caption{Performance comparison of ablation study.}
\label{tab:ablation}
\end{table}

\subsection{Ablation Study}
In this section, we conduct some ablation studies to investigate the effectiveness of three important components of our proposed model: (1) Explicit relations; (2) Implicit similarities; (3) The product layer. We set four comparative settings, and the performances of them have been shown in Table \ref{tab:ablation}. The details of the four settings are listed below:
\begin{itemize}
    \item \textbf{RER} (Remove Explicit Relations) does not consider explicit relations between questions and skills, i.e. removes $\mathcal{L}_1(\pmb{Q},\pmb{S})$ from Eqn.(\ref{eq:optimization}).
    \item \textbf{RIS} (Remove Implicit Similarities) does not consider implicit similarities among questions and skills, i.e. removes $\mathcal{L}_2(\pmb{Q})$ and $\mathcal{L}_3(\pmb{S})$ from Eqn.(\ref{eq:optimization}).
    \item \textbf{RPL} (Remove Product Layer) directly concatenates $\pmb{q}$, $\pmb{s}'$ and $\pmb{ a}$ as the pre-trained question embedding instead of using product layer.
    \item \textbf{RPF} (Replace Product Layer with Fully Connected Layer) concatenates $\pmb{q}$, $\pmb{s}'$ and $\pmb{ a}$ as the input of a fully connected layer instead of product layer.
\end{itemize}

Except for the changes mentioned above, the other parts of the models and experimental settings remain identical. 
\begin{figure}[ht]
    \includegraphics[width=0.48\textwidth]{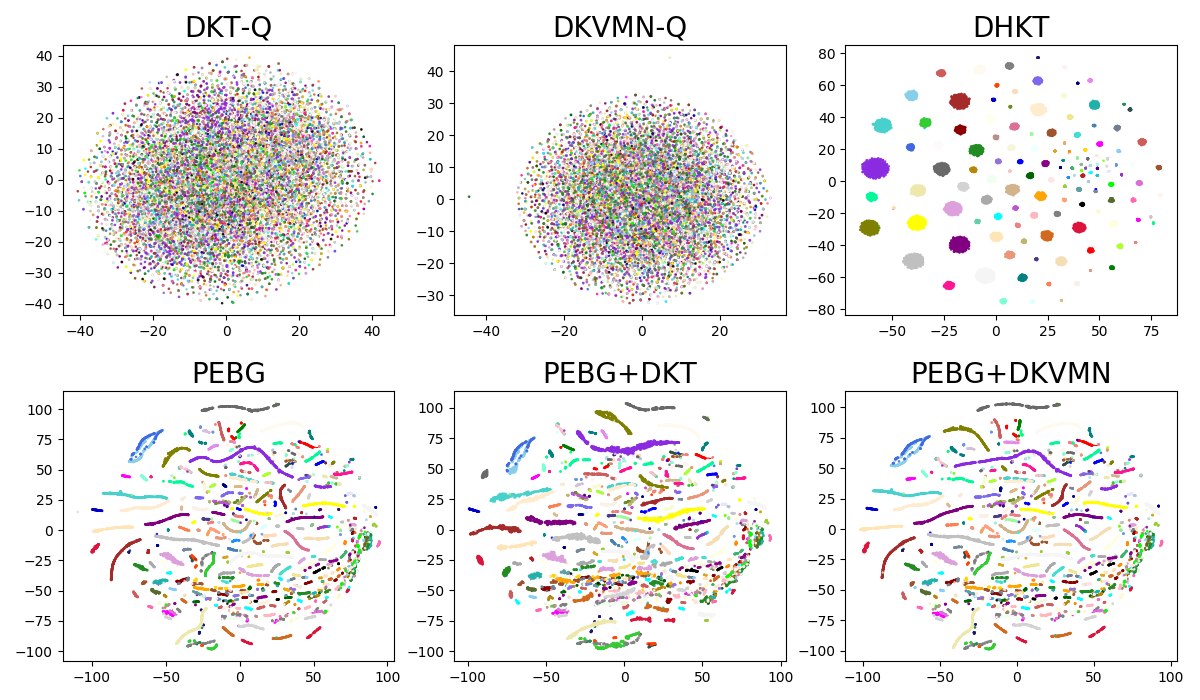}
    \caption{Comparison of question embeddings learned by question-level deep KT models on the ASSIST09 dataset. The questions related to the same skill are labeled in the same color.}
    \label{fig:embed}
\end{figure}

From Table \ref{tab:ablation} we can find that (1) PEBG+DKT and PEBG+DKVMN perform best indicates the efficacy of different components of the models. (2) The models show a similar degree of decline when removing explicit relations and implicit similarities, which means these two pieces of information are equally important. (3) Removing the product layer hurts the performance badly, and using a fully connected layer also has a lower performance. By exploration of feature interactions, the product layer is promising to learn high-order latent patterns compared to directly concatenating features. (4) Without the product layer, RPF and RPL are standard graph embedding methods, which use the first-order and second-order neighbor information of the bipartite graph. And our proposed pre-trained model PEBG can better improve the performance of existing deep KT models.

\subsection{Embedding Comparison}
We use t-SNE \cite{maaten2008visualizing} to project the multi-dimensional question embeddings pre-trained by PEBG and question embeddings learned by other question-level deep KT models to the 2-D points. 

Figure \ref{fig:embed} shows the visualization of question embeddings. Question embeddings learned by DKT and DKVMN are randomly mixed, which completely loses the relations among questions and skills. Question embeddings of different skills learned by DHKT are completely separated, which fails to capture implicit similarities.Question embeddings pre-trained by PEBG are well structured. Questions in the same skill are close to each other, and questions that do not relate to common skills are well separated. PEBG+DKT and PEBG+DKVMN fine-tune the question embeddings pre-trained by PEBG to make them more suitable for the KT task while retaining the relations among questions and skills.

\section{Conclusion}

In this paper, we propose a novel pre-training model PEBG, 
which first formulates the question-skill relations as a bipartite graph and introduce a product layer to learn low-dimensional question embeddings for knowledge tracing. Experiments on real-world datasets show that PEBG significantly improves the performance of existing deep KT models. Besides, visualization study shows the effectiveness of PEBG to capture question embeddings, which provides an intuitive explanation of its high performance.

\section*{Acknowledgements}
The corresponding author Yong Yu thanks the support of NSFC (61702327 61772333).

\bibliographystyle{named}
\bibliography{ijcai20}

\end{document}